# Effect of Hexagonal Boron Nitride on Energy Band Gap of Graphene Antidot Structures

Penchalaiah Palla[*]   and   J.P.Raina

Center for Nanotechnology Research, Vellore Institute of Technology, Vellore -632014, Tamil Nadu, India.
* E-mail of the corresponding author: penchalaiah.palla@vit.ac.in

*The authors are grateful to the Chancellor and Management of V.I.T. University for generous support in funding the Advanced Computing facility and for providing specialized computational tools, which enabled us to bring out the work reported here.*

**Abstract**

The zero band gap ($E_g$) graphene becomes narrow $E_g$ semiconductor when graphene is patterned with periodic array of hexagonal shaped antidots, the resultant is the hexagonal Graphene Antidot Lattice (hGAL). Based on the number of atomic chains between antidots, hGALs can be even and odd. The even hGALs (ehGAL) are narrow $E_g$ semiconductors and odd hGALs (ohGAL) are semi-metals. The $E_g$ opening up by hGALs is not sufficient to operate a realistic switching transistor. Also hGAL transistors realized on $Si/SiO_2$ substrate are suffering with low carrier mobility and ON-OFF current ratio. In order to achieve a sizable $E_g$ with good mobility, AB Bernal stacked hGALs on hexagonal Boron Nitride (hBN), ABA Bernal stacked hBN / hGAL / hBN sandwiched structures and AB misaligned hGAL /hBN structures are reported here for the first time. Using the first principles method the electronic structure calculations are performed. A sizable $E_g$ of about 1.04 eV (940+100 meV ) is opened when smallest neck width medium radius ehGAL supported on hBN and about 1.1 eV (940 + 200 meV ) is opened when the same is sandwiched between hBN layers. A band gap on the order of 71 meV is opened for Bernal stacked ohGAL / hBN and nearly 142 meV opened for hBN / ohGAL /hBN structures for smallest radius and width of nine atomic chains between antidots. Unlike a misaligned graphene on hBN, the misaligned ohGAL/hBN structure shows increased $E_g$. This study could open up new ways of band gap engineering for graphene based nanostructures.

## 1. Introduction

Graphene, an atomic thick two dimensional material, is one of the extraordinary materials studied so far in nanoscience and nanotechnology. However, the intrinsic electronic structure of graphene limits its direct application for Graphene Field Effect Transistor (GFET) devices, which is due to graphene's lack of energy band gap (Schwierz F. 2010). Several strategies have been developed and experimentally verified, for opening band gap of graphene. One such method is the Graphene Antidot Lattices (GALs) (Pedersen T. G. et al. 2008). In its basic form the GAL can be viewed as a graphene sheet with a periodic nanometer scale antidots (holes).The various aspects of GALs have been studied theoretically, for e.g., electronic properties (Furst, J. A. et al. 2009) (Vanevic, M. et al. 2009), optical properties (Pedersen, T. G., Flindt, C., Pedersen, J., Jauho,A.P. et al. 2008), magnetic properties (Yu, D.C. et al. 2008), electron-phonon coupling (Vukmirovic, N. et al.2010) (Stojanovic, V. M. et al. 2010), band gap scaling (Liu, W. et al. 2009) (Zhang, A. H. et al. 2011), chemical functionalization with different species (Ouyang, F. P. et al. 2010), thermoelectric properties (Gunst, T. et al. 2011) and transport properties ( Lopata,K. et al. 2010) (Jippo, H. et al 2011). Many experimental techniques for fabrication of GALs have been developed to verify the interesting quantum mechanical effects predicted by theory. Graphene films with antidot diameters of 20-150 nm, spacing between the antidots of less than10 nm and cell sizes of 35- 400 nm have been fabricated via electron beam lithography, nano imprint lithography, block copolymer lithography, and self-assembling of monodispersed colloidal microspheres ( Begliarbekov, M. et al. 2011) (Bai, J. W. et al. 2010) (Liang, X. et al. 2010).

The calculations have predicted that antidot lattices change the electrical properties of graphene from semi-metallic to semiconducting, where the opened energy band gap can be tuned by the size, shape, and symmetry of both the antidot and the lattice unit cell (Pedersen T. G. et al. 2008) (Jippo, H. et al 2011). The induced band gap in GALs is approximately proportional to the antidot diameter and inversely proportional to the super lattice cell area, so one can get a band gap value of ~200 meV for a unit cell of 10 nm (Pedersen T. G. et al. 2008). Some experimental studies have demonstrated that GALs have an effective energy gap of 100 meV and Field effect Transistors made from GALs (GA-FETs) have an ON - OFF current ratio of up to 10, which suggests the promising behavior for the scheme of GALs (Kim, M. et al. 2010) (Sinitskii, A. et al. 2010 ).

After examination of the published theoretical works of GALs in the literature, we noticed that most of the early works on GALs have focused on single layer GALs with various size, shape, and symmetry of both the antidot and the lattice unit cell/s with or without $SiO_2$/Si substrate. Similarly most of the experimental papers concentrated on GALs supported on $SiO_2$/ Si or sandwiched between $SiO_2$ and $SiO_2$/ Si layers. As mentioned above with only single layer GALs or GALs supported on $SiO_2$/ Si, we can achieve a band gap of ~200 meV, which is not sufficient to provide OFF-state for GA-FETs at room temperature without leakage. To operate switched FETs at room temperature band gap of ~500 meV and above is needed. Therefore, it is further necessary to increase sizable band gap of GALs to make use of in nanoelectronics applications. Moreover, GA-FET realized on Si/$SiO_2$ substrate suffer from low carrier mobility due to impurities, rough surface, charged surface states and presence of surface optical phonons on $SiO_2$ (Bai, J. W. et al. 2010 ) (Chen, J.-H. et al .2008) (Ponomarenko, L. A. et al. 2009 ). Therefore, it is highly desirable to develop an effective strategy to open a tunable and sizable band gap for GALs without significant loss of carrier mobility and ON-state current. However, to fabricate any effective GA-FET device, it should be supported on a suitable substrate. Recently, graphene has been transferred to hexagonal boron nitride (hBN) substrate. Experimental studies found that graphene on hBN substrate yields much higher mobility, which is comparable to that of the suspended graphene. Moreover, the mobility is one order of magnitude larger than that of graphene supported on $SiO_2$ because the atomically flat h-BN substrate is free from charge impurities and dangling bonds (Dean, C. R. et al. 2010) ( Xue,J. et al. 2011). Theoretical studies have predicted that there exists a possibility of inducing a band gap in graphene when supported on the hBN substrate (Giovannetti, G.et al. 2007 ) but, recent experimental investigations find no band gap in this bilayer system (Dean, C. R. et al. 2010) ( Xue,J. et al. 2011). The first principle many-body calculations with GW approximation (Kharche, N. et al. 2011 ) also confirm the fact that slight misalignment of graphene with respect to hBN (possible in experiments), closes the band gap induced by hBN in Bernal stacked graphene.

To the best of our knowledge, so far, there are no studies reported on the effect of hBN on the GALs. Therefore, first-principles calculations are required for accurate prediction of electronic structure of this novel material system. In order to further tune to achieve sizable band gap with good mobility, a combination of GAL and hBN is likely to be one of the future development directions for high-performance GA-FET devices. To this end, GAL structures should be sandwiched between hBN layers.

## 2. Model and Methodology

In this paper, we examine one particular GAL (with tri-angular lattice symmetry) in detail as a demonstration. We have chosen hexagonal shaped unit cell with hexagonal shaped antidot with zigzag antidot edges. Based on the number of atomic chains between antidots (we call it neck-width), the hGALs can be classified as odd hGALs (ohGALs) and even hGALs (ehGALs). In this study we consider three different types of bilayer and tri layer structures, the perfectly AB Bernal stacked hGALs over hBN ( hGAL / hBN), perfectly ABA Bernal stacked hGAL sandwiched between two hBN single layers ( hBN / hGAL / hBN), and misaligned hGAL over hBN layer, to investigate the band gap variation with odd and even neck-width and radii. We found unique properties for odd and even neck width hGALs under the influence of hBN.

Recently, Clar Sextet theory on GALs (Petersen, R. et al. 2011) suggested that triangular array of ehGALs are more stable and technologically realizable. The recent paper on GALs by Ouyang et al., designed a new ohGAL with zero band gap, which is same as pristine graphene (Ouyang, F. et al. 2011). The recent experimental papers on, graphene supported on hBN suggested that hBN substrate is the ideal choice for graphene to improve the mobility of the carriers in graphene (Dean, C. R. et al. 2010) ( Xue,J. et al. 2011).  Furthermore, the band gap of GALs can be tuned by controlling the lattice parameters such as neck-width and antidot size (Pedersen T. G. et al. 2008). If one can consider those above issues then the question arises, what is the effect of hBN on the electronic structure of hGALs? To address this issue, we first considered (i) ehGALs, (ii) ohGALs without hBN substrate, (iii) both perfectly AB Bernal stacked ehGALs / hBN and ohGALs / hBN, (iv) both perfectly ABA Bernal stacked   hBN / ehGAL / hBN and hBN / ohGAL / hBN as both substrate or gate dielectric for bottom gate and top gate dielectric(i.e. hGAL is sandwiched between hBN layers), and finally (v) misaligned ohGAL / hBN. These studies are being reported first time to the best of our knowledge.

The geometry of GAL studied in this paper is illustrated in Figure 1a and 1b. Each system is designated by unit cell {R, W}, where 'R' is the hole radius. R is described as a unit in terms of the number of removed hexagonal carbon chains. Therefore, R is an integer and is related to the number of removed carbon atoms as $N_{removed} = 6R^2$ as reported (Ouyang, F. et al. 2011).  'W' is the neck width or spacing between antidots, described in terms of the number of atomic chains between the antidots. The antidots considered, in this study, form a triangular lattice and have a hexagonal shape (see Figure 1a). The unit cell of hGALs (Figure 1) is characterized by the side length 'L' the number of the outermost carbon atoms on each edge-side (L = 3 in Figure 1a). To expand or reduce the unit cell, one usually adds the extra outermost zigzag carbon chains or remove chains outside the unit cell. This increases or decreases L by 1, and thus L is an integer.    However, when the assembled lattice with even-width unit cells are inspected, it can be seen that the neck width W between the nearest-neighbor antidots changes incrementally by 2. Similarly, for odd unit cells, when the assembled lattice with odd-unit cells are inspected, it can be seen that the neck width W between the nearest-neighbor antidots changes incrementally by 1(Ouyang, F. et al. 2011).

Here ATK 11.8.2, a semi-empirical and first principle simulation package, is used to compute the optimized geometries, to find equilibrium distance between hGAL and hBN and perform electronic structure calculations. The geometry optimization is performed for both the atomic positions and the interlayer equilibrium distance, until the maximum force on each atom is less than 0.05 eV/A°. The optimized geometries is calculated using ATK-DFTB (Slater-Koster model) (Stokbro K. et al. 2010) using cp2K parameters with a grid mesh cutoff of 20 Hartree. The Total energy, Band structure and Density of states (DOS) calculations are performed in the framework of density functional theory (DFT) within the local density approximation (LDA) using double-zeta-polarized (DZP) basis set as implemented in the ATK-DFT(Brandbyge M.et al.2002) package. The norm-conserving pseudo-potentials and the PZ parameterization for the exchange-correlation functionals are used with grid mesh cutoff of 75 Hartree. To ensure negligible interaction between periodic images, a large value (10 Å) of the vacuum region is used. The Brillouin zone is sampled using Monkhorst_Pack grids of different sizes depending on the size of the unit cell. The k-point sampling is varied for different sized super cells. The ranges of k-point sampling included in the calculation are 36×36×1, 13×13×1, 9×9×1, 6×6×1, and 3×3×1.

In constructing the supercell for composite hGAL layer on top of hBN system, we use the most stable configuration with one carbon over Boron (B), and the other carbon centered above a hBN hexagon hole-site, i.e., hGAL is perfectly AB Bernal stacked on hBN layer by an angle of 60º (Giovannetti, G.et al. 2007).   We also consider that hGAL and hBN are commensurate and lattice mismatch between the hGAL and hBN is <2%. After optimizing the system with DFTB and performing DFT within LDA, the minimum energy configuration corresponds to separation of hGALs and hBN layers of 3.1 Å, Figure 2b, which is reasonably comparable to reported theoretical value of 3.21 Å for hBN / Graphene/ hBN (Quhe, R.et al. 2012). Similarly, in constructing the composite perfectly ABA Bernal stacked hBN / hGAL / hBN sandwiched structure, we use the most stable ABA configuration as mentioned in the reference (Quhe, R.et al. 2012), with 3.1 Å as interlayer spacing on either side of the hGAL. Depending on the size of the supercell, the various values of lattice constants with a=b, c=16 Å (for bilayer) and c= 20 Å (for trilayer) are considered for constructing the super cell.

## 3. Result and Discussion

The atomic schematics, the corresponding band structures and density of states (DOS) of the hGALs with even and odd neck widths W without hBN are illustrated in Figure 1, for fixed holes of R = 1. The gray and white balls represent carbon and hydrogen atoms (after passivation), respectively. A {1, 4} hexagonal unit cell (Unit cell and antidot enclosed by black hexagonal line) characterized by a side length L = 3 and a hole radius R = 1 and its assembly into a lattice with an even neck width of W = 4 is shown in Figure 1a. The corresponding DOS and band structure of ehGAL of size {1, 4} is shown in Figure 1c and 1d. With even W, the patterning of antidots opened a substantial band gap around the Fermi level. For example, the band gap of the {1, 4} hGAL is 940 meV (Figure 1d) and this decreases as the unit cell becomes larger. These results are consistent with previous studies (Furst, J. A. et al. 2009). The origin of the band gap opening can be related to the chiral symmetry breaking, or inter-valley mixing (Lee, S.H. et al. 2011). Recent first principle studies on the origin of the band gap of ehGALs reveal that, quantum confinement effects are not the origin of the band gap of hGALs, since band gap is varying with super cell-size (Lee, S.H. et al. 2011). The atomic schematic of ohGAL of size {1,7} with unit cell highlighted by black color (Figure 1b), the corresponding DOS and band structure with odd neck width W=7 without hBN are illustrated in Figure 1e and 1f, for fixed hole size of R = 1. However, with odd W, the hGALs exhibit semi-metallic behavior like pristine graphene as shown in Figure 1f. So, hGALs with odd W have zero bandgap, which is consistent with previous study (Ouyang, F. et al. 2011).

The effect of hBN on odd width structures are demonstrated in comparison with band structures and DOS of ohGAL without hBN, perfectly AB Bernal stacked ohGAL / hBN and perfectly ABA Bernal stacked hBN / ohGAL / hBN as shown in Figure 2. For example, the atomic schematic of ohGAL of size {1, 7}, without hBN is shown in Figure 2a. The top and side views of perfectly AB Bernal stacked {1, 7} ohGAL with hBN is shown in Figure 2b. The top and side views of perfectly ABA Bernal stacked {1, 7} ohGAL with hBN is shown in Figure 2c. Figure 2d is the corresponding band structure comparison of structures shown from Figure 2a to 2c. Figure 2e is the enlarged and the highlighted portion of dotted rectangle of Figure 2d. Figure 2f is the corresponding DOS comparison of structures shown from Figure 2a to Figure 2c. Odd width hGAL without hBN of size {1, 7} exhibits a band gap of 0.0 meV. Interestingly, the band gap of hGALs of odd width increases to 65.84 meV due to the substrate induced lattice symmetry breaking (Giovannetti, G.et al. 2007) when it is supported on hBN. And when ohGAL is sandwiched between hBN layers, the band gap further increases to 126.3 meV that is an increment of 60.46 meV when compared with ohGAL/hBN as shown in Figure 2e. The change of ohGAL dispersion from linear to parabolic with hBN interaction indicates a slight decrease in mobility of ohGAL due to increase in band gap, as shown in figure 2e.

The effect of hBN on even width structures is demonstrated in comparison with band structures and DOS of ehGAL without hBN, perfectly AB Bernal stacked ehGAL / hBN and perfectly ABA Bernal stacked hBN / ehGAL / hBN, as shown in Figure 3. For e.g., the atomic schematic of ehGAL {1, 4, without hBN is shown in Figure 3a. The top and side views of perfectly AB Bernal stacked {1, 4} ehGAL with hBN is shown in Figure 3b. The top and side views of perfectly ABA Bernal stacked {1, 4} ehGAL with hBN is shown in Figure 3c. Figure 3d is the corresponding band structure comparison of structures shown from Figure 3a to 3c. Figure 3e is the enlarged and the highlighted portion of dotted rectangle of Figure 3d. Figure 3f is the corresponding DOS comparison of structures shown from Figure 3a to 3c with enlarged portion as inset. Even width hGALs without hBN, for example, of size {1, 4} exhibits a band gap of 937.9 meV. Contrary to ohGAL structures, when small radius ehGAL structure is supported on hBN, for example, the band gap of {1, 4} decreases to 923 meV, which is nearly 15 meV less. And when {1, 4} is sandwiched between hBN layers, the band gap further decreases to 917.6 meV i.e. by 5.4 meV and further decreases by 20.3 meV when compared with {1, 4} without hBN, as shown in figure 3(e).The decrease in band gap in both cases of perfectly AB Bernal stacked ehGAL / hBN and perfectly ABA Bernal stacked hBN / ehGAL / hBN can be understood because of polarization effects similar to graphone / hBN and hBN / graphone / hBN (Kharche, N. et al. 2011).

The band gaps of odd and even hGALs with various W and R are systematically calculated and analyzed under the influence of hBN as shown in Figure 4. As shown in Figure 4a, the variation of band gap as a function of neck width W for fixed R1 (one hexagon ring or 6 carbon atoms removed to make hexagonal antidot) and R2 (2 hexagonal rings or 24 carbon atoms removed to make antidot) are plotted for ohGAL without hBN, perfectly AB Bernal stacked ohGAL / hBN and perfectly ABA Bernal stacked hBN /ohGAL / hBN sandwiched structures. For ohGAL supported on hBN, the band gap opens and band gap further increases by increasing the spacing W between the antidots. Compared to perfectly AB Bernal stacked ohGAL / hBN, the perfectly ABA Bernal stacked hBN / ohGAL/ hBN sandwiched structures show more band gap and it further increases by increasing the spacing between the antidots. A band gap on the order of 71 meV is opened for Bernal stacked ohGAL / hBN and nearly 142 meV opened for hBN / ohGAL /hBN structures for smallest radius and width of nine atomic chains between antidots. The variation of band gap as a function of antidot radius R by keeping neck width W = 3 constant is plotted for ohGAL without hBN, perfectly AB Bernal stacked ohGAL / hBN and perfectly ABA Bernal stacked hBN / ohGAL / hBN structures as shown in Figure 4b. For ohGAL without hBN, the band gap remains zero when R increases. However, for ohGAL / hBN and hBN / ohGAL/ hBN structures the band gap decreases exponentially and becomes almost zero when R = 6. The band gap of hBN / ohGAL/ hBN sandwiched structure is more than the band gap of ohGAL/ hBN when R is smaller and the band gap difference between two structures decreases when R increases. The similar behavior is noticed for other odd widths with some sample test points (other points not shown here, which involve heavy computation time). In contrast to even widths, for odd width structures, when widths (for example is W = 5, W = 3, W= 1) decrease, the band gap also decreases and curves follow similar shape as W=3 with increasing R.

In contrast to band gap variation of odd width structures, for even width structures the band gap decreases with increasing spacing between antidots. The variation of band gap as a function of neck width W for fixed R1 and R2 are plotted for ehGAL without hBN, perfectly AB Bernal stacked ehGAL/hBN and perfectly ABA Bernal stacked hBN / ehGAL/ hBN sandwiched structures, as shown in Figure 4c. There is a band gap for ehGALs without hBN substrate. For ehGAL supported on hBN, the band gap decreases and it further decreases by increasing the spacing between the antidots. Compared to ehGAL/hBN, the ehGAL sandwiched between hBN shows less band gap and band gap further decreases by increasing the spacing between the antidots. For smaller R1, W = 2 set of curves, show higher band gap than R2, however, for beyond W= 2, the trend changes and R2 set of curves show more band gap than R1 and the difference in band gap values between R1 and R2 sets remain constant. In the case of ehGALs the effect of addition of hBN layers does not cause much variation in band gap values for both R1 and R2. This is in contrast to the behaviour of ohGAL/hBN and hBN/ohGAL/hBN.

The variation of band gap as a function of antidot radius R, while keeping neck width W constant, is plotted for ehGAL, perfectly AB Bernal stacked ehGAL/hBN and perfectly ABA Bernal stacked hBN / ehGAL/ hBN, as shown in Figure 4d. There are three sets of curves as shown for W=2, W=4 and W=6, respectively. Each set consists of 3 curves one is for ehGAL, second one for ehGAL/hBN and third one for hBN / ehGAL/ hBN sandwiched structure. For smaller R and W values, for e.g. R=1, and W=2, the band gap is high and is almost same (~1800 meV ) for ehGAL, ehGAL/hBN and hBN/ehGAL/hBN. For W=2, R > 1, the addition of hBN introduces differences in band gap values and that difference increases for ehGAL/hBN and hBN/ehGAL/hBN structures when compared with ehGAL. For W=2, R>1, ehGAL / hBN structures induced (added) an extra amount of about 100 meV to the band gap of ehGAL without hBN. Similarly the sandwiched hBN / ehGAL / hBN structures induced (added) an extra amount of 200 meV band gap to the band gap of ehGAL without hBN. Therefore, the total band gap is the sum of the band gap of ehGAL plus band gap due to hBN( ~100 meV for bilayer and ~200 meV for trilayer). As the spacing between antidots increases, the effect of chiral symmetry breaking changes and band gap decreases as shown in figure 4d. Moreover, as R increases, i.e. beyond R = 4, band gap decreases and the band gap difference between various widths (W=2, 4, 6) also decreases. The above calculations show that the effect of hBN is more on smaller W and higher R ehGALs. And hBN can reliably induce an extra band gap in ehGALs with perfect Bernal stacking.

The perfect Bernal stacking of hGAL on hBN may be difficult to obtain in the experiments and random orientation is more probable similar to graphene on hBN (Dean, C. R. et al. 2010) ( Xue,J. et al. 2011). To mimic the random orientation, we simulate a small sized ohGAL supercell (to reduce computation time) of size {1, 1} where the stacking between the ohGAL layer and the underlying hBN substrate deviates from the ideal Bernal stacking. In the perfect Bernal stacking, the ohGAL layer is rotated by an angle θ = 60º with respect to hBN. We consider a rotation angle of 21.8º.   The commensurate supercell for this rotation contain 28 atoms (8 C, 6 H, 7 B, 7 N). The supercell for 21.8º rotation is depicted in the Figure 5a. The commensuration conditions derived in reference (Laissardiere, G.T.D. et al. 2010) are used to build the supercell.

In this supercell, the sublattice asymmetry induced by the hBN substrate on the hGAL layer is significantly increased compared to the Bernal stacking. The band structure of the misaligned ohGAL (θ =21.8º) is compared with that of the perfect Bernal stacked ohGAL and misaligned graphene as shown in Figure 5b. The enlarged view of highlighted dotted rectangle of figure 5b is shown in figure 5c.  Unlike in the Perfectly Bernal stacked graphene (which has finite band gap and then gap closes due to the slight misalignment), the perfect Bernal stacked ohGAL has a finite band gap of 27.5 meV, which increases with slight misalignment to 95.86 meV . The Moire patterns larger than those shown in Figure 5a were recently observed in graphene supported on hBN ( Xue, J. et al. 2011). These graphene samples indeed showed zero band gaps similar to our result. The above calculations indicate that the hBN substrate can reliably induce a band gap in hGALs with perfect Bernal stacking and even in misalignment.

**4. Conclusion**

In summary, our first principles Density functional calculations suggest that ohGALs are semimetals with zero bandgap like pristine graphene, while ehGALs are  narrow band semiconductors. When odd width ohGAL is AB Bernal stacked on hBN substrate band gap opens up due to sub-lattice asymmetry in ohGAL induced by hBN and band gap further increases when ohGAL is ABA Bernal stacked like  hBN  /hGAL / hBN . For e.g., {1, 9} odd width ohGAL is AB Bernal stacked on hBN substrate, a band gap of the order of 71 meV opens up, When {1, 9} ohGAL is sandwiched between hBN layers in ABA stacking, the band gap further increases to 142 meV. For this ohGALs, when antidot radius R increases, the band gap tends to decrease. In the case of ehGALs the effect of addition of hBN layers as ehGAL/hBN and hBN / ehGAL/hBN   does not cause much variation in band gap values for  both R1 and R2  for increasing antidot width. This is in contrast to the behaviour of ohGAL/hBN and hBN/ohGAL/hBN with increasing antidot width.  For ohGAL without hBN, the band gap remains zero when R increases. However, for ohGAL / hBN and hBN / ohGAL/ hBN structures the band gap decreases exponentially and becomes almost zero when R = 6. The band gap of hBN / ohGAL/ hBN sandwiched structure is more than the band gap of  ohGAL/ hBN when R is smaller and the band gap difference between two structures decreases when R increases. An extra amount of above 100 meV   is (induced) added to the band gap of  single layer smallest width and higher R   ehGAL is AB Bernal stacked on hBN. Similarly, an extra amount of more than 200 meV   is added to the band gap of single layer smallest width and higher R   ehGAL is sandwiched   between hBN layers (in ABA stacking). In contrast for ehGAL, for R values greater than 4, the behavior of band gap is similar to W=2 for all higher values of W.   Unlike in the Perfect Bernal stacked graphene (which has finite band gap, which closes due to the slight mis-orientation), the perfect Bernal stacked {1, 1} ohGAL has a finite band gap of 27.5 meV, which increases to 95.86 meV with 21.8º mis-orientation. One can achieve higher band gap values, when smallest neck width and higher radius ehGALs are sandwiched between hBN layers in ABA Bernal stacking. Similarly to achieve higher band gap, smallest radius and higher neck width ohGALs should be sandwiched between hBN layers in ABA stacking.   The results indicate that the hBN substrate can reliably induce a band gap in hGALs with perfect Bernal stacking even when misalignment occurs. Thus, calculated band structure and DOS suggest that for field effect devices employing hGAL, as a channel material, hBN can be used as a substrate as well as a dielectric layer separating the hGAL channel and the gate electrode(s).   The demonstration of sizable band gap opening using hBN in this work provides theoretical direction for further experimental verification.

**Figure 1.** Atomic schematic of hexagonal unit cell of size (a) {1, 4} (Unit cell and antidot highlighted by black hexagonal line) characterized by a side length L = 3 and a hole radius R = 1 and its assembly into a lattice with an even neck width of W= 4. (b) {1, 7} (Unit cell highlighted by black color) and its assembly into a lattice with an odd neck width of W= 7. (c) The DOS of unit cell {1, 4}. (d) The band structure of {1, 4}. (e) The DOS of unit cell {1,7}. (f) The band structure of unit cell {1, 7}.

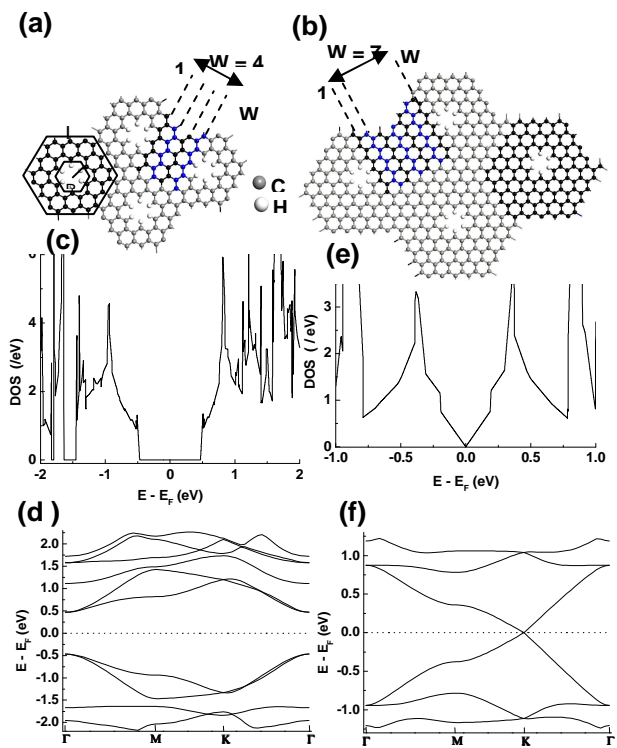

**Figure 2.** (a) Single layer odd hGAL unit cell of size {1, 7} without hBN. (b)Top and side views of AB Bernal stacked {1, 7} hGAL / hBN. (c) Top and side views of ABA Bernal stacked hBN / {1, 7} hGAL / hBN sandwiched structure. (d) band structure comparison of structures (a)-(c). (e) enlarged and highlighted portion of dotted rectangle of (d). (f) DOS comparison of structures (a) -(c).

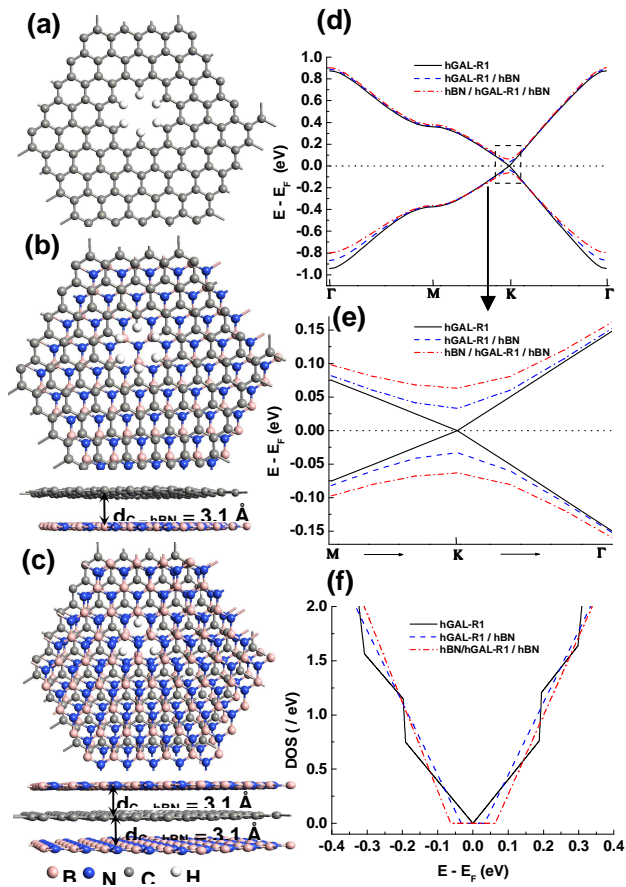

**Figure 3.** (a) Single layer even hGAL unit cell of size {1, 4} without hBN. (b)Top and side views of AB Bernal stacked {1, 4} hGAL / hBN. (c) top and side views of ABA Bernal stacked hBN / {1, 7} hGAL / hBN sandwiched structure. (d) band structure comparison of structures (a)-(c). (e) the enlarged and highlighted portion of dotted rectangle of (d). (f) DOS comparison of structures (a)-(c). The inset shows the enlarged and highlighted portion of dotted rectangle of (f).

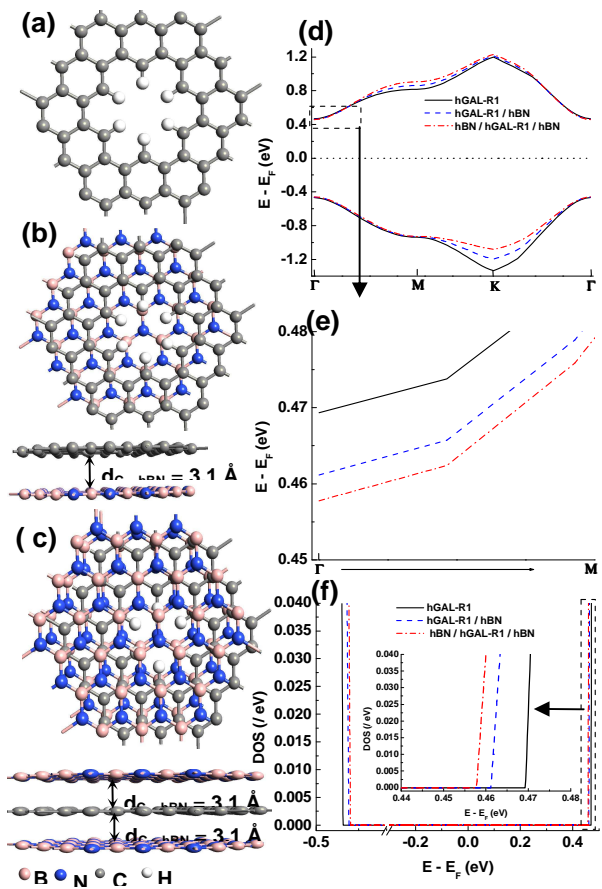

**Figure 4.** (a) The band gap as a function of odd neck width W with fixed R = 1 and R = 2 for ohGAL, ohGAL / hBN and hBN /ohGAL / hBN structures. (b) The bandgap as a function of R for odd neck width W = 3 for ohGAL, ohGAL / hBN and hBN / ohGAL / hBN structures. (c) The band gap as a function of even neck width W with fixed R = 1 and R = 2 for ehGAL, ehGAL / hBN and hBN / ehGAL / hBN structures. (d) The bandgap as a function of R for constant even neck width W=2, W=4 and W= 6 values for ehGAL, ehGAL / hBN and hBN / ehGAL / hBN structures.

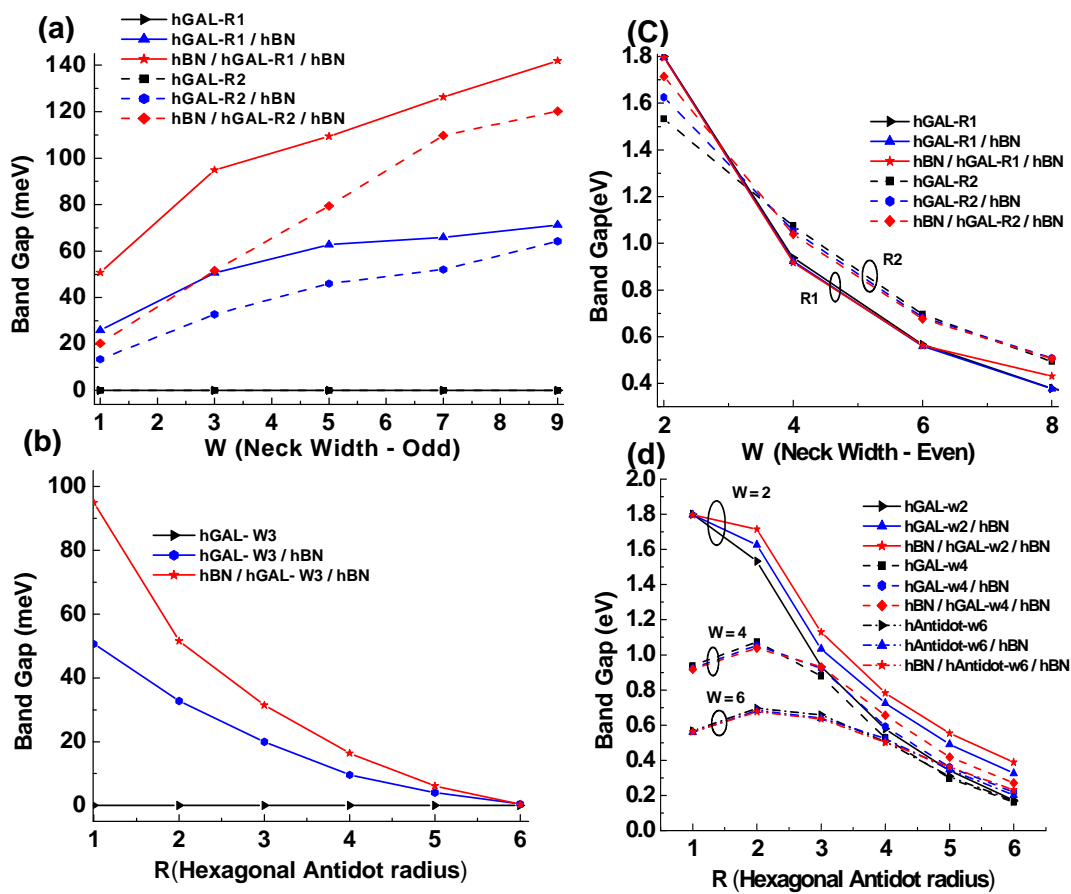

**Figure 5**. (a) Atomic schematic of a commensurate unit cell of the misaligned {1, 1} ohGAL / hBN. (b) The band structures of Bernal stacked ohGAL on hBN substrate, misaligned ohGAL / hBN and misaligned graphene / hBN . (c) The enlarged view of dotted rectangle shown in (b).

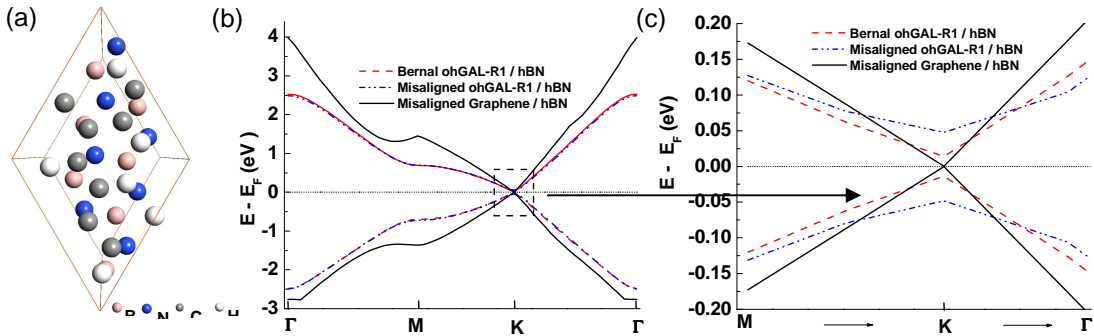